\documentclass[proceedings]{JHEP} 

\usepackage{epsfig}                     

\newbox\mybox
\newcommand\fverb{\setbox\mybox=\hbox\bgroup\verb}
\newcommand\fverbdo{\egroup\medskip\noindent\fbox{\unhbox\mybox}\ }
\newcommand\fverbit{\egroup\item[\fbox{\unhbox\mybox}]}

\font\beeg=cmr17 scaled 1600            
\newcommand\init[1]{\setbox\mybox=\hbox{{\beeg #1}~}%
                   \noindent\global\hangindent=\wd\mybox\global\hangafter-2%
                   \sc\smash{\llap {\lower 13.2pt \box\mybox}}}

\title{Mott dissociation of D-mesons at the chiral phase transition and 
anomalous J/$\psi$ suppression}

\author{D.\ Blaschke, G.\ Burau, Yu. L.\ Kalinovsky\\
        $^{db,gb}$ Fachbereich Physik, Rostock University, D-18051 Rostock, 
        Germany\\
        $^{yulk}$Laboratory for Computing Techniques and Automation, JINR, 
        141980 Dubna, Russia\\
        (presented by Yu.L. Kalinovsky)\\
        Email: \email{david@darss.mpg.uni-rostock.de}}

\conference{Heavy Quark Physics 5, Dubna, Russia, 6-8 April 2000}

\abstract{We investigate the in-medium modification of the charmonium breakup 
processes due to the Mott-effect for D-mesons at the chiral phase 
transition.
A model calculation for the process J/$\psi+\pi\to D+\bar D^* + h.c.$ 
is presented which demonstrates that threshold effects in the thermal averaged
breakup cross section can be explained as a Mott transition where final 
state quark-antiquark bound states enter the continuum of resonant states at 
the QCD phase transition.
Applications to heavy-ion collisions within a modified Glauber 
model scenario and the phenomenon of anomalous J/$\psi$ suppression in the 
CERN NA50 experiment are addressed.}


\begin{document} 

\section{Introduction} 
Recent results of the CERN NA50 collaboration on anomalous J/$\psi$ 
suppression \cite{na50} in ultrarelativistic Pb-Pb collisions at 158 AGeV 
have renewed the quest for an explanation of the processes which may cause the 
rather sudden drop of the J/$\psi$ production cross section for transverse 
energies above $E_T\sim 40$ GeV in this experiment.
An effect like this was predicted as a signal for quark gluon plasma 
formation \cite{ms} due to screening of the $c\bar c$ interaction.
Soon after that it became clear that for temperatures and densities just above
the deconfinement transition the Mott effect for the J/$\psi$ does not occur
and that a kinetic process as e.g. dissociation by meson or quark impact is
required to dissolve the J/$\psi$ \cite{b}.

In this paper, we suggest that at the chiral/deconfinement phase transition 
the charmonium breakup reaction cross sections are critically enhanced 
since the open charm states of the dissociation processes become unbound 
(Mott effect) so that the reaction thresholds are effectively lowered.
We present a model calculation for the particular process 
J/$\psi+\pi\to D+\bar D^* + h.c.$ in a hot pion gas in order to demonstrate 
that the Mott dissociation of the D-mesons at the chiral phase transition
leads to a threshold effect for the J/$\psi$ suppression ratio when applied
to the case of Pb-Pb collisions at CERN within a modified Glauber model. 

\section{In-medium modification of charmonium break-up cross sections}

The inverse lifetime of a charmonium state in a hot and dense many-
particle system is given by the imaginary part of the selfenergy 
$\tau^{-1}(p)=\Gamma(p)= \Sigma^>(p) \mp \Sigma^<(p)$. 
\FIGURE{\epsfig{file=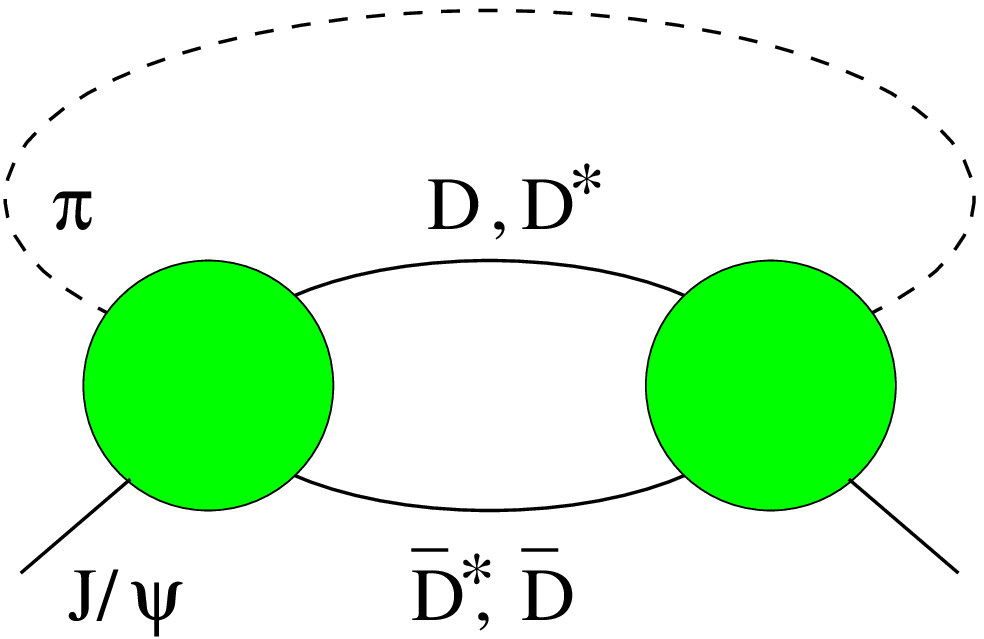,width=7.0cm}
\caption{Diagrammatic representation of the complex selfenergy
for the J/$\psi$ due to breakup in (of-shell) $D,\bar D*$ pairs by impact of
pions from a hot medium.}
\label{fig:cex}}
In the Born collision approximation for the dominant process in a hot pion gas,
as shown in Fig. \ref{fig:cex}, we have \cite{kb}
\begin{eqnarray}
\Sigma^{\stackrel{>}{<}}(p,\omega) &=& \int_{p'} 
\int_{p_1}\int_{p_2} (2\pi)^4 \delta(p+p'-p_1-p_2)\nonumber\\
&\times&\left|{\cal M}\right|^2 
G^{\stackrel{<}{>}}_\pi(p')~ 
G^{\stackrel{>}{<}}_{D_1}(p_1)~ 
G^{\stackrel{>}{<}}_{D_2}(p_2)~, \nonumber
\end{eqnarray}
where the thermal Green functions
$G^{>}_{i}(p)=[1 \pm f_i(p)]A_i(p)$ and $G^{<}_{i}(p)=f_i(p)A_i(p)$ 
are defined by the spectral function $A_i(p)$ and the distribution function
$f_i(p)$ of the state $i$; $\int_p=\int \frac{d^4p}{(2\pi)^4}$.
In the low density approximation for the final states ($f_{D_i}(p) \approx 0$),
one can safely neglect $\Sigma^<(p)$ so that 
\begin{eqnarray}
\tau^{-1}(p)&=& \int_{p'} 
\int_{p_2}\int_{p_2} 
(2\pi)^4 \delta(p+p'-p_1-p_2) \nonumber \\
&\times&\left| {\cal M}\right|^2 
f_\pi(p')~ A_\pi(p')~ A_{D_1}(p_1)~ A_{D_2}(p_2).
\nonumber
\end{eqnarray}
With the differential cross section 
\begin{displaymath}
\frac{d\sigma}{dt} = \frac{1}{16\pi} 
\frac{\left| {\cal M}(s,t)\right|^2}
{\lambda(s,M_\psi^2,s')}~, 
\end{displaymath}
using $s=(p+p')^2, t=(p-p_1)^2, s'=p'^2$ and 
$\sqrt{\lambda(s,M_\psi^2,s')}=
2v_{\rm rel}\sqrt{{\bf p}^2+M_\psi^2}\sqrt{{\bf p'}^2+s'}$
one can show that the $J/\psi$ relaxation time in a hot 
pion gas is given by
\begin{eqnarray}
\label{tau}
\tau^{-1}(p)= \int \frac{d^3{\bf p'}}{(2\pi)^3} 
\int ds'&& f_\pi({\bf p'},s')~ A_\pi(s')\nonumber\\ 
&\times&v_{\rm rel}~ \sigma^*(s)~, 
\end{eqnarray}
where the pion spectral function decides whether the medium consists of 
resonant (off-shell) correlations or true $q\bar q$ bound states. 
The in-medium breakup cross section is given by
\begin{equation}
\label{sig*}
\sigma^*(s) = \int
ds_1~ ds_2~ 
 A_{D_1}(s_1)~ A_{D_2}(s_2)~ \sigma(s;s_1,s_2)~. 
\end{equation}
Note that there are two kinds of medium effects due to (i) the spectral 
functions of the final states and (ii) the explicit medium dependence of the 
matrix element ${\cal M}$. 
In the following model calculation we will use the approximation 
$\sigma(s;s_1,s_2)\approx\sigma^{\rm vac}(s;s_1,s_2)$ 
justified by the locality of the transition matrix ${\cal M}$ which makes it 
rather inert against medium influence.
 
\section{Model calculation}

The quark exchange processes in meson-meson scattering can be calculated 
within the diagrammatic approach of Barnes and Swanson \cite{bs} which allows 
a generalization to finite temperatures in the thermodynamic Green function
technique \cite{br}. 
This technique has been applied to the calculation of J/$\psi$ break-up cross
sections by pion impact in \cite{mbq}, where also a fit formula has been 
given.
The approach has been extended to excited charmonia states and consideration 
of rho-meson impact recently \cite{wsb}.
The generic form of the resulting cross section (given a band of uncertainty)
can be fit to the form 
\begin{equation}
\label{sig0}
\sigma^{\rm vac}(s;M_{D_1}^2,M_{D_2}^2)=
\sigma_0 \ln(s/s_0) \exp(s/\lambda^2)~,
\end{equation}   
where $s_0=(M_{D_1}+M_{D_2})^2$ is the threshold for the process to occur,
$\sigma_0=7.5\cdot 10^{9}$ mb and $\lambda=0.9$ GeV.
An alternative fit is given in \cite{bsw}.

Recently, the charmonium dissociation processes have been calculated also in
an effective Lagrangian approach \cite{mm,haglin}, but this method 
introduces new phenomenological parameters and ignores the quark substructure.
The development of a unifying approach on the basis of a relativistic confining
quark model is in progress \cite{kb99}.
Detailed numerical comparison cannot be made at present.

The major modification of the charmonium break-up process which we expect at 
finite temperatures in a hot meson gas comes from the Mott-effect for the light
as well as the heavy mesons. 
At finite temperatures when the chiral symmetry in the light quark sector is 
restored, the continuum threshold for light-heavy quark pairs drops below the 
mass of the D-mesons so that they are no longer bound states constrained
to their mass shell, but become rather broad resonant correlations in the 
continuum. 
This Mott-effect has been discussed within relativistic quark models 
\cite{b+93} for the
light meson sector but can also be generalized to the case of heavy mesons 
\cite{gk}. Applying a confining quark model \cite{kb99} we have obtained the 
critical temperatures $T^{\rm Mott}_{D^*}=110$ MeV, 
$T^{\rm Mott}_{D}=140$ MeV. 

In order to study the implications of the D-meson Mott effect for the 
charmonium breakup we adopt here a Breit-Wigner form
for the spectral functions 
\begin{eqnarray}
\label{ad}
A_i(s)&=&
\frac{1}{\pi}
\frac{\Gamma_i(T)~M_i(T)}{(s-M_i^2(T))^2+\Gamma_i^2(T)M_i^2(T)}
~,
\end{eqnarray} 
which in the limit of vanishing width $\Gamma_i(T)\to 0$ goes over into the 
delta function $\delta(s-M_i^2)$ for a bound state in the channel $i$.
The width of the D-mesons shall be modeled by a microscopic approach.
For our exploratory calculation, we adopt here
\begin{equation}
\Gamma_D(T)=c (T-T^{\rm Mott}_D)\Theta(T-T^{\rm Mott}_D)~,
\end{equation}
where the coefficient $c=2.67$ is assumed to be universal for the pseudoscalar 
D-mesons and it is obtained from a fit to the pion width above the pion Mott 
temperature, see \cite{b+95}. For the D-meson mass we have 
$M_D(T)=M_D+0.75\Gamma_D(T)$.
The result for the in-medium J/$\psi$ break-up cross section (\ref{sig*}) is 
shown in Fig. \ref{fig:sig_t}.

\FIGURE{\epsfig{file=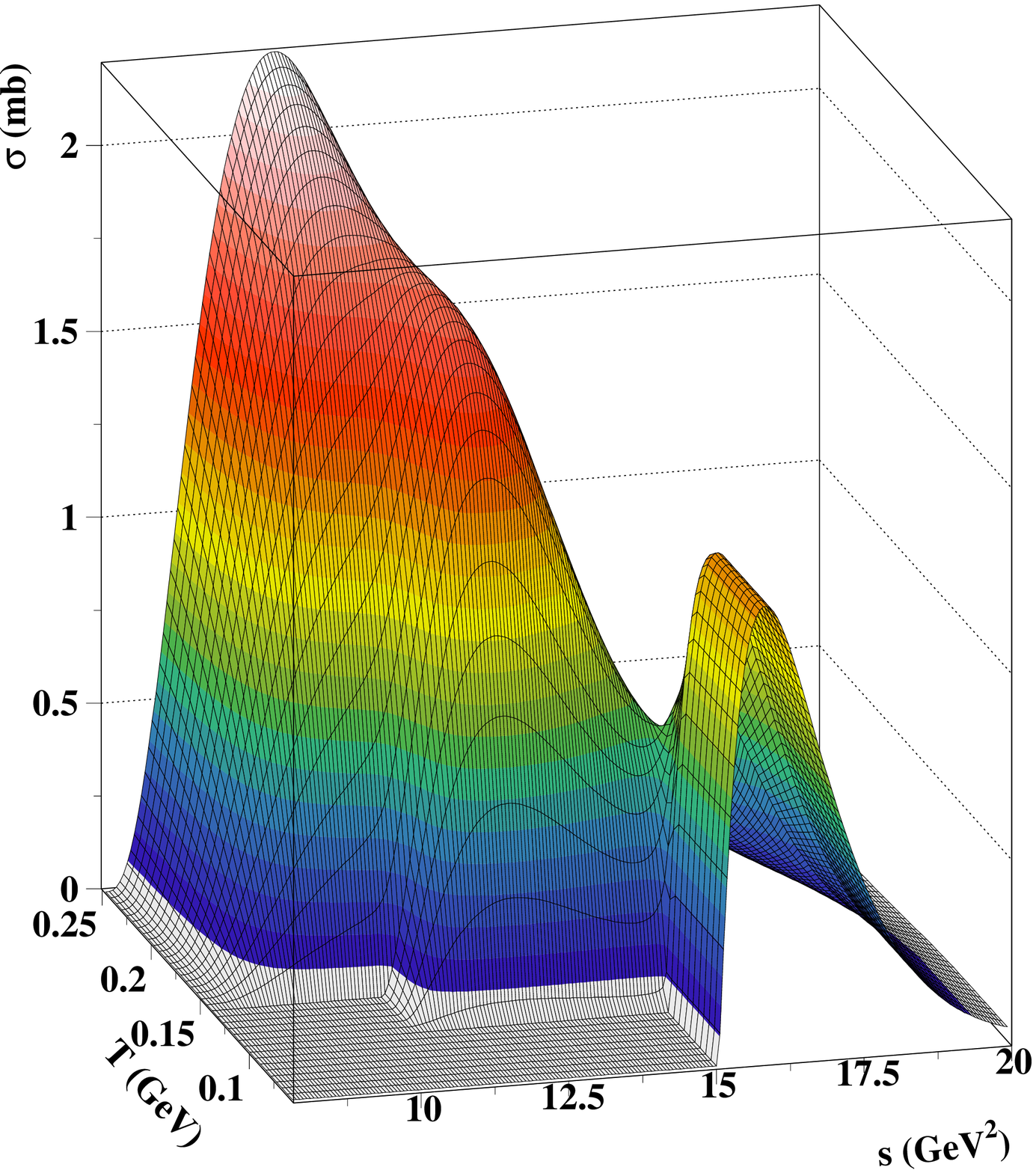,width=7.0cm}
\caption{Energy- and temperature dependent in-medium 
J/$\psi$ break-up cross section for pion impact. Thresholds occur at the Mott 
temperatures for the open charm mesons: $T^{\rm Mott}_{D^*}=110$ MeV, 
$T^{\rm Mott}_{D}=140$ MeV}
\label{fig:sig_t}}

With $M_{D^*}=2.01$ GeV and $M_{\bar D}=1.87$ GeV follows for the threshold
$s_0=15.05$ GeV$^2$. At a temperature $T=140$ MeV, where the D-meson can still
be considered as a true bound state, the $D^*$ meson has already entered the
continuum and is a resonance with a half width of about 80 MeV.
Due to the Mott effect for the open charm mesons, the charmonium dissociation 
processes become "subthreshold" ones and their cross sections which are peaked 
at threshold rise and spread to lower onset with cms energy. 
This is expected to enhance the rate for the charmonium dissociation processes
in a hot meson gas. 

\section{J/$\psi$ dissociation in a hot pion gas}

We calculate the inverse relaxation time 
for a J/$\psi$ at rest in a hot pion gas according to Eq. (\ref{tau}),
\begin{eqnarray}
\tau^{-1}(T)
&=&\int\frac{d^3{\bf p}}{(2\pi)^3}
f_\pi({\bf p}) \frac{|{\bf p}|}{E_\pi({\bf p})} \sigma^*(s) \nonumber\\
&=&<\sigma^* v_{\rm rel}>n_\pi(T)~,\nonumber
\end{eqnarray}
where we have assumed on-shell pions with the dispersion relation 
$E_\pi({\bf p})=\sqrt{{\bf p}^2+M_\pi^2}$ obeying the thermal Bose distribution
function $f_\pi({\bf p})=3[\exp[(E_\pi({\bf p})-\mu)/T]-1]^{-1}$, $n_\pi(T)$ 
being the pion density. The cms energy of the pion impact on a J/$\psi$
at rest is $s({\bf p})=M_\pi^2+M_\psi^2+2M_\psi E_\pi({\bf p})$.
The result for the temperature dependence of the thermal averaged J/$\psi$ 
breakup cross section $<\sigma^* v_{\rm rel}>$ is shown in Fig. \ref{fig:sigv}.

\FIGURE{\epsfig{file=sigv.eps,width=7cm}
\caption{Temperature dependence of the in-medium J/$\psi$ break-up cross 
section for different quark masses.}
\label{fig:sigv}}

This quantity has to be compared to the nuclear absorption cross section for 
the J/$\psi$ of about 3 mb which has been extracted from charmonium suppression
data in p-A collisions \cite{hhk}.
It is remarkable that it is practically negligible below the D-meson Mott
temperature $T^{\rm Mott}_{D^*}=110$ MeV but comparable to the nuclear 
absorption cross section above the chiral/deconfinement temperature of
$T_{\rm crit}\approx 150$ MeV. 
Therefore we expect the in-medium enhanced charmonium 
dissociation processes to be sufficiently effective to destroy the 
charmonium state on its way through the hot fireball of the heavy-ion collision
and to provide an explanation of the observed anomalous J/$\psi$ suppression 
phenomenon \cite{na50}.

\section{Applications to Pb-Pb collisions at CERN}

A more detailed comparison with the recent data from the NA50 collaboration
requires a model for the heavy ion collision. In the following we will use
a modified Glauber model \cite{wong,mb} which takes into accout the formation 
of secondaries ("soft matter") during the collision and the propagation of the 
charmonium state through the debris of the collision.
The suppression function is defined by 
\begin{eqnarray}
S(E_T) &=& S_N(E_T)\exp\left[-\int_{t_0}^{t_f} dt ~\tau^{-1}(n(t))\right] 
\nonumber \\
&=&  S_N(E_T)\exp\left[\int_{n_0(E_T)}^{n_f} dn <\sigma^* v_{\rm rel}>\right] 
,\nonumber
\end{eqnarray}
where we have assumed a a nuclear absorption systematics 
$S_N(E_T)=18+36~\exp(-0.26\sqrt{E_T})$ and longitudinal expansion according 
to $n(t)=n_0(E_T)t_0/t$. 
Within a Glauber model fit to the NA50 experiment \cite{mb} we obtain the 
$n_0(E_T)$ as a parameter form depending on the impact parameter $b$, where 
$E_T(b)/{\rm MeV}=130 - b/{\rm fm} $ and 
$n_0(b)/{\rm fm}^{-3} = 1.2\sqrt{1-(b/10.8~{\rm fm})^2}$.  

\FIGURE{\epsfig{file=pbpbexp.eps,width=7.0cm}
\caption{$E_T$ dependence of the J/$\psi$ suppression ratio without (dashed) 
and with (solid, long-dashed) in-medium break-up process compared to NA50 
data.}
\label{fig:pbpbexp}}

In Fig. \ref{fig:pbpbexp} we show the result for our model calculation in 
comparison with the data for Pb-Pb collision experiments of the NA50 
collaboration \cite{na50}. The temperature (pion density) dependence of the 
J/$\psi$ breakup cross section which exhibits a critical enhancement at the
D-meson Mott temperature, see Fig. \ref{fig:sigv} leads to a threshold 
behaviour in the $E_T$ dependence of the J/$\psi$ suppression ratio.
This is in qualitative agreement with the data and the mechanism for the 
occurrence of thresholds in this quantity can be considered as a possible 
explanation for the observed deviation from the ordinary nuclear absorption
systematics (anomalous J/$\psi$ suppression) in the NA50 experiment.

\section{Summary and Outlook}

In this contribution we have presented an approach to charmonium breakup in a
hot and dense medium which is applicable in the vicinity of the 
chiral/deconfinement phase transition where mesonic bound states get dissolved
in a Mott-type transition and should be described as resonant correlations in
the quark plasma instead. This description can be achieved using the concept 
of the spectral function which can be obtained from relativistic quark models
in a systematic way. The result of an exploratory calculation employing a
temperature-dependent Breit-Wigner spectral function for open charm mesons
presented in this paper has demonstrated that heavy-flavor dissociation 
processes are critically enhanced at the QCD phase transition and could lead 
in the charmonium sector to the phenomenon of anomalous J/$\psi$ suppression.

In subsequant work we will relax systematically approximations which have been 
made in the one presented here and improve inputs which have been used. 
In particular, we 
will investigate the off-shell behaviour of the charmonium breakup cross 
section in the vacuum (\ref{sig0}) and calculate the spectral function 
(\ref{ad}) at finite temperature from a relativistic quark model.
Dyson-Schwinger equations provide a nonperturbative, field-\-theo\-retical 
approach which has recently been applied also to heavy-meson observables 
\cite{ikr}
and have proven successful for finite-temperature generalization \cite{bbkr}. 
Further intermediate open charm states can be considered, the states in the
dense environment should include rho mesons and nucleons besides of the 
pions which all can be treated as off-shell quark correlations at the QCD 
phase transition. 

In future experiments at LHC the charm distribution in the created fireball 
may be not negligible so that the approximation $f_{D_i}(p)\approx 0$ has to 
be relaxed. In this case, not only the gain process 
($D\bar D$ annihilation \cite{ko,pbm}) encoded in the $\Sigma^{<}$ has to be
included but at the same time the Bose enhancement factors in the $G^{>}_{i}$
functions have to be considered.
  
We want to emphasize that this approach to the kinetics of open charm and 
charmonium states in a dense medium provides a framework for a systematic 
investigation of processes relevant to QCD plasma diagnostics in heavy-ion 
collisions using  heavy quark bound states.

\section{Acknowledgements}

This work has been supported by the Heisenberg-Landau program for 
scientific collaborations between Germany and the JINR Dubna and by 
the DFG Graduiertenkolleg ``Stark korrelierte 
Vielteilchensysteme'' at the University of Rostock. 
We thank T. Barnes, J. H\"ufner, M.A. Ivanov,
V.N. Pervushin, C.D. Roberts, G. R\"opke and P.C. Tandy for their discussions
and stimulating interest in our work.
D.B. acknowledges the hospitality of the INT Seattle where this work has been 
started, and of the ECT* Trento where it has been completed.


\begin{thebibliography}{999}
\bibitem{na50}
M.C. Abreu et al., NA50 Collaboration,
\plb {477} {2000} {28}.
\bibitem{ms}
T. Matsui, H. Satz, \plb {178} {1986} {416}.
\bibitem{b}
D. Blaschke, \npa {525} {1991} {296c}.
\bibitem{kb} 
L.P. Kadanoff and G. Baym, {\em Quantum Statistical Mechanics},
Addison-Wesley, 1962.
\bibitem{bs}
T. Barnes, E.S. Swanson, \prd {46} {1992} {131}.
\bibitem{br}
D. Blaschke, G. R\"opke, \plb {299} {1993} {332}.
\bibitem{mbq} 
K. Martins, D. Blaschke, E. Quack, \prc {51} {1995} {2723}.
\bibitem{wsb}
C.-Y. Wong, E.S. Swanson, T. Barnes, 
{\it Cross Sections for pi- and rho-induced Dissociation of J/psi and psi'},
\hepph{9912431}.
\bibitem{bsw}
T. Barnes, E.S. Swanson, C.-Y. Wong, {\em Charmonium + Light Hadron Cross 
Sections}, these proceedings, \nuclth{0006012}.
\bibitem{mm}
S.G. Matinyan, B. M\"uller, \prc {58} {1998} {2994}.
\bibitem{haglin}
K.L. Haglin, \prc {61} {2000} {031902}.
\bibitem{kb99}
D. Blaschke, G. Burau, M.A. Ivanov, Yu.L. Kalinovsky and P.C. Tandy,
in {\em Progress in Nonequilibrium Green's Functions}, World Scientific,
Singapore (2000), Ed. M. Bonitz, p. 392.
\bibitem{b+93}
D. Blaschke et al., {\em Z. Phys.} {\bf A 346} {1993} {85}.
\bibitem{gk}
F.O. Gottfried and S.P. Klevansky, \plb {286} {1992} {221}.
\bibitem{b+95}
D. Blaschke et al., \npa {592} {1995} {561}.
\bibitem{hhk}
Y.B. He, J. H\"ufner, B.Z. Kopeliovich, \plb {477} {2000} 93.
\bibitem{wong}
C.-Y. Wong, \prl {76} {1996} {196}.
\bibitem{mb}
K. Martins, D. Blaschke, \hepph{9802250}.
\bibitem{ikr}
M.A. Ivanov, Yu.L. Kalinovsky, C.D. Roberts, {\em Heavy-Meson Observables via 
Dyson-Schwinger Equations}, these proceedings, \hepph{0006189}.
\bibitem{bbkr}
A. Bender, D. Blaschke, Yu.L. Kalinovsky, C.D. Roberts, \prl {77}{1996}{3724}.
\bibitem{ko}
C.M. Ko, X.N. Wang, B. Zhang, X.F. Zhang, \plb {444} {1998} {237}.
\bibitem{pbm}
P. Braun-Munzinger, K. Redlich, \npa {661} {1999} 546.
\end{thebibliography}
\end{document}